\documentclass[10pt]{iopart}
\usepackage{iopams}
\usepackage{graphicx}
\usepackage{cite}
\usepackage[pdfstartview=FitH, pdfpagemode=None, pdfpagelayout=OneColumn, colorlinks=true, linkcolor=blue, citecolor = blue]{hyperref}
\newcommand{\text}[1]{\mathrm{#1}}
\newcommand{\eqref}[1]{(\ref{#1})}

\begin{document}
 \title[Comparison of AWAKE simulations and measurements]{Proton beam defocusing in AWAKE: comparison of simulations and measurements}

\author{A.A.~Gorn$^{1,2}$,
M.~Turner$^{3}$,
E.~Adli$^{4}$,
R.~Agnello$^{5}$,
M.~Aladi$^{6}$,
Y.~Andrebe$^{5}$,
O.~Apsimon$^{7,8}$,
R.~Apsimon$^{7,8}$,
A.-M.~Bachmann$^{3,9,10}$,
M.A.~Baistrukov$^{1,2}$,
F.~Batsch$^{3,9,10}$,
M.~Bergamaschi$^{3}$,
P.~Blanchard$^{5}$,
P.N.~Burrows$^{11}$,
B.~Buttensch{\"o}n$^{12}$,
A.~Caldwell$^{9}$,
J.~Chappell$^{13}$,
E.~Chevallay$^{3}$,
M.~Chung$^{14}$,
D.A.~Cooke$^{13}$,
H.~Damerau$^{3}$,
C.~Davut$^{7,15}$,
G.~Demeter$^{6}$,
L.H.~Deubner$^{16}$,
A.~Dexter$^{7,8}$,
G.P.~Djotyan$^{6}$,
S.~Doebert$^{3}$,
J.~Farmer$^{3,9}$,
A.~Fasoli$^{5}$,
V.N.~Fedosseev$^{3}$,
R.~Fiorito$^{7,17}$,
R.A.~Fonseca$^{18,19}$,
F.~Friebel$^{3}$,
I.~Furno$^{5}$,
L.~Garolfi$^{20}$,
S.~Gessner$^{3,21}$,
B.~Goddard$^{3}$,
I.~Gorgisyan$^{3}$,
E.~Granados$^{3}$,
M.~Granetzny$^{22}$,
O.~Grulke$^{12,23}$,
E.~Gschwendtner$^{3}$,
V.~Hafych$^{9}$,
A.~Hartin$^{13}$,
A.~Helm$^{19}$,
J.R.~Henderson$^{7,24}$,
A.~Howling$^{5}$,
M.~H{\"u}ther$^{9}$,
R.~Jacquier$^{5}$,
I.Yu.~Kargapolov$^{1,2}$,
M.{\'A}.~Kedves$^{6}$,
F.~Keeble$^{13}$,
M.D.~Kelisani$^{3}$,
S.-Y.~Kim$^{14}$,
F.~Kraus$^{16}$,
M.~Krupa$^{3}$,
T.~Lefevre$^{3}$,
L.~Liang$^{7,15}$,
S.~Liu$^{20}$,
N.~Lopes$^{19}$,
K.V.~Lotov$^{1,2}$,
M.~Martyanov$^{9}$,
S.~Mazzoni$^{3}$,
D.~Medina~Godoy$^{3}$,
V.A.~Minakov$^{1,2}$,
J.T.~Moody$^{9}$,
P.I.~Morales~Guzm\'{a}n$^{9}$,
M.~Moreira$^{3,19}$,
T.~Nechaeva$^{25}$,
H.~Panuganti$^{3}$,
A.~Pardons$^{3}$,
F.~Pe\~na~Asmus$^{9,10}$,
A.~Perera$^{7,17}$,
A.~Petrenko$^{1}$,
J.~Pucek$^{9}$,
A.~Pukhov$^{26}$,
B.~R\'{a}czkevi$^{6}$,
R.L.~Ramjiawan$^{3,11}$,
S.~Rey$^{3}$,
H.~Ruhl$^{27}$,
H.~Saberi$^{3}$,
O.~Schmitz$^{22}$,
E.~Senes$^{3,11}$,
P.~Sherwood$^{13}$,
L.O.~Silva$^{19}$,
R.I.~Spitsyn$^{1,2}$,
P.V.~Tuev$^{1,2}$,
F.~Velotti$^{3}$,
L.~Verra$^{3,9,10}$,
V.A.~Verzilov$^{20}$,
J.~Vieira$^{19}$,
C.P.~Welsch$^{7,17}$,
B.~Williamson$^{7,15}$,
M.~Wing$^{13}$,
J.~Wolfenden$^{7,17}$,
B.~Woolley$^{3}$,
G.~Xia$^{7,15}$,
M.~Zepp$^{22}$,
G.~Zevi~Della~Porta$^{3}$ (The AWAKE Collaboration)}
\address{$^{1}$ Budker Institute of Nuclear Physics SB RAS, Novosibirsk, Russia}
\address{$^{2}$ Novosibirsk State University, Novosibirsk, Russia}
\address{$^{3}$ CERN, Geneva, Switzerland}
\address{$^{4}$ University of Oslo, Oslo, Norway}
\address{$^{5}$ Ecole Polytechnique Federale de Lausanne (EPFL), Swiss Plasma Center (SPC), Lausanne, Switzerland}
\address{$^{6}$ Wigner Research Center for Physics, Budapest, Hungary}
\address{$^{7}$ Cockcroft Institute, Daresbury, UK}
\address{$^{8}$ Lancaster University, Lancaster, UK}
\address{$^{9}$ Max Planck Institute for Physics, Munich, Germany}
\address{$^{10}$ Technical University Munich, Munich, Germany}
\address{$^{11}$ John Adams Institute, Oxford University, Oxford, UK}
\address{$^{12}$ Max Planck Institute for Plasma Physics, Greifswald, Germany}
\address{$^{13}$ UCL, London, UK}
\address{$^{14}$ UNIST, Ulsan, Republic of Korea}
\address{$^{15}$ University of Manchester, Manchester, UK}
\address{$^{16}$ Philipps-Universit{\"a}t Marburg, Marburg, Germany}
\address{$^{17}$ University of Liverpool, Liverpool, UK}
\address{$^{18}$ ISCTE - Instituto Universit\'{e}ario de Lisboa, Portugal}
\address{$^{19}$ GoLP/Instituto de Plasmas e Fus\~{a}o Nuclear, Instituto Superior T\'{e}cnico, Universidade de Lisboa, Lisbon, Portugal}
\address{$^{20}$ TRIUMF, Vancouver, Canada}
\address{$^{21}$ SLAC National Accelerator Laboratory, Menlo Park, CA}
\address{$^{22}$ University of Wisconsin, Madison, Wisconsin, USA}
\address{$^{23}$ Technical University of Denmark, Lyngby, Denmark}
\address{$^{24}$ Accelerator Science and Technology Centre, ASTeC, STFC Daresbury Laboratory, Warrington, UK}
\address{$^{25}$ Belarusian State University, 220030 Minsk, Belarus}
\address{$^{26}$ Heinrich-Heine-Universit{\"a}t D{\"u}sseldorf, D{\"u}sseldorf, Germany}
\address{$^{27}$ Ludwig-Maximilians-Universit{\"a}t, Munich, Germany}

\vspace{10pt}
\begin{indented}
\item[]\today
\end{indented}

\begin{abstract}
In 2017, AWAKE demonstrated the seeded self-modulation (SSM) of a 400 GeV proton beam from the Super Proton Synchrotron (SPS) at CERN. The angular distribution of the protons deflected due to SSM is a quantitative measure of the process, which agrees with simulations by the two-dimensional (axisymmetric) particle-in-cell code LCODE. Agreement is achieved for beam populations between $10^{11}$ and $3 \times 10^{11}$ particles, various plasma density gradients ($-20 \div 20\%$) and two plasma densities ($2\times 10^{14} \text{cm}^{-3}$ and  $7 \times 10^{14} \text{cm}^{-3}$). The agreement is reached only in the case of a wide enough simulation box (at least five plasma wavelengths).
\end{abstract}


\ioptwocol
\sloppy
\section{Introduction}
\label{s1}

Acceleration of particles in plasmas, or plasma wakefield acceleration, offers the possibility to reach multi-GeV and, potentially, TeV range electron and positron energies in facilities that are orders of magnitude smaller than modern high-energy accelerators \cite{Nat.445-741,Nat.524-442,PPCF58-034017, PRL122-084801,RMP81-1229,RAST9-19,RAST9-63,NatPhys5-363,PoP18-103101,RAST9-85}.
Studies of novel accelerators significantly benefit from numerical simulations that go in parallel with them \cite{RAST9-165}. These simulations complement fragmentary experimental data \cite{RMP90-035002} and form a complete picture of physical processes that occur at tiny spatial and temporal scales inside the plasma and, therefore, cannot be measured.

Recently, the Advanced WAKefield Experiment (AWAKE) at CERN \cite{NIMA-829-3,NIMA-829-76,PPCF60-014046} demonstrated seeded self-modulation of a long proton beam in the plasma \cite{PRL122-054801,PRL122-054802} and electron acceleration by the wakefield of this beam \cite{Nat.561-363}. This milestone achievement opens the way for using proton beams from modern synchrotrons as drivers for plasma wakefield acceleration \cite{RAST9-85,PPCF56-084013,EPJC76-463}. Because of the high proton energy, the micro-bunching and acceleration occurs in a single plasma cell, avoiding difficulties related to staged acceleration \cite{PRST-AB3-071301,PRST-AB15-111303,PRL116-124801,IPAC16-2561}.

In this paper, we show that simulations of beam self-modulation agree with related AWAKE measurements. The axisymmetric self-modulation process is thus well understood in the sense that simulations include the most important effects, they can help in understanding details of the process and can serve as a valuable starting point for future predictions. We also discuss how to conduct these simulations properly, and show what the results would look like otherwise. The AWAKE data used for the studies reported in this paper were taken during the 2017 and early 2018 running periods.

\section{Methods}

\begin{figure*}[t]
\includegraphics{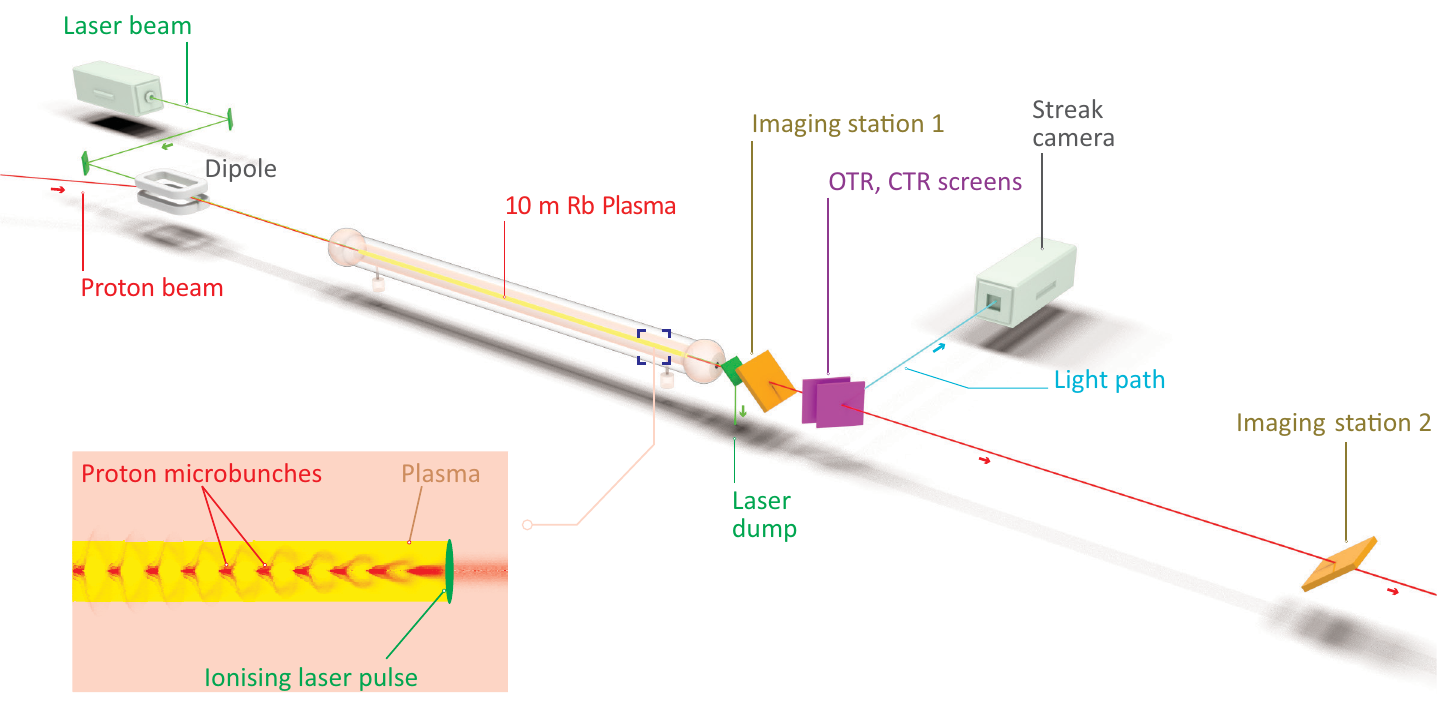}
\caption{Setup of AWAKE experiments on beam self-modulation.}\label{fig1}
\end{figure*}

AWAKE uses the proton beam from the SPS \cite{IPAC17-1747}. The proton beam co-propagates with a short laser pulse through a 10-meter-long cylindrical cell filled with rubidium vapor (Fig.\,\ref{fig1}). The laser pulse quickly ionizes the vapor, so that the newly appearing plasma interacts only with the rear part of the beam, as if the beam would be cut at the pulse position. The leading edge of this ``sharply cut'' beam seeds a small amplitude wakefield in the plasma that grows in space and time and eventually converts the beam into a train of short micro-bunches \cite{PRL104-255003,PoP22-103110}. The self-modulation occurs due to defocusing of protons from the space between the micro-bunches. The radial momentum gained by the defocused protons is an integral quantitative measure of the self-modulation process: the stronger the longitudinal wakefield micro-bunch formation, the larger the transverse momentum of the defocused protons \cite{PRL122-054801,NIMA-829-314,NAPAC16-707}.

\begin{figure}[tb]
\includegraphics[width=\columnwidth]{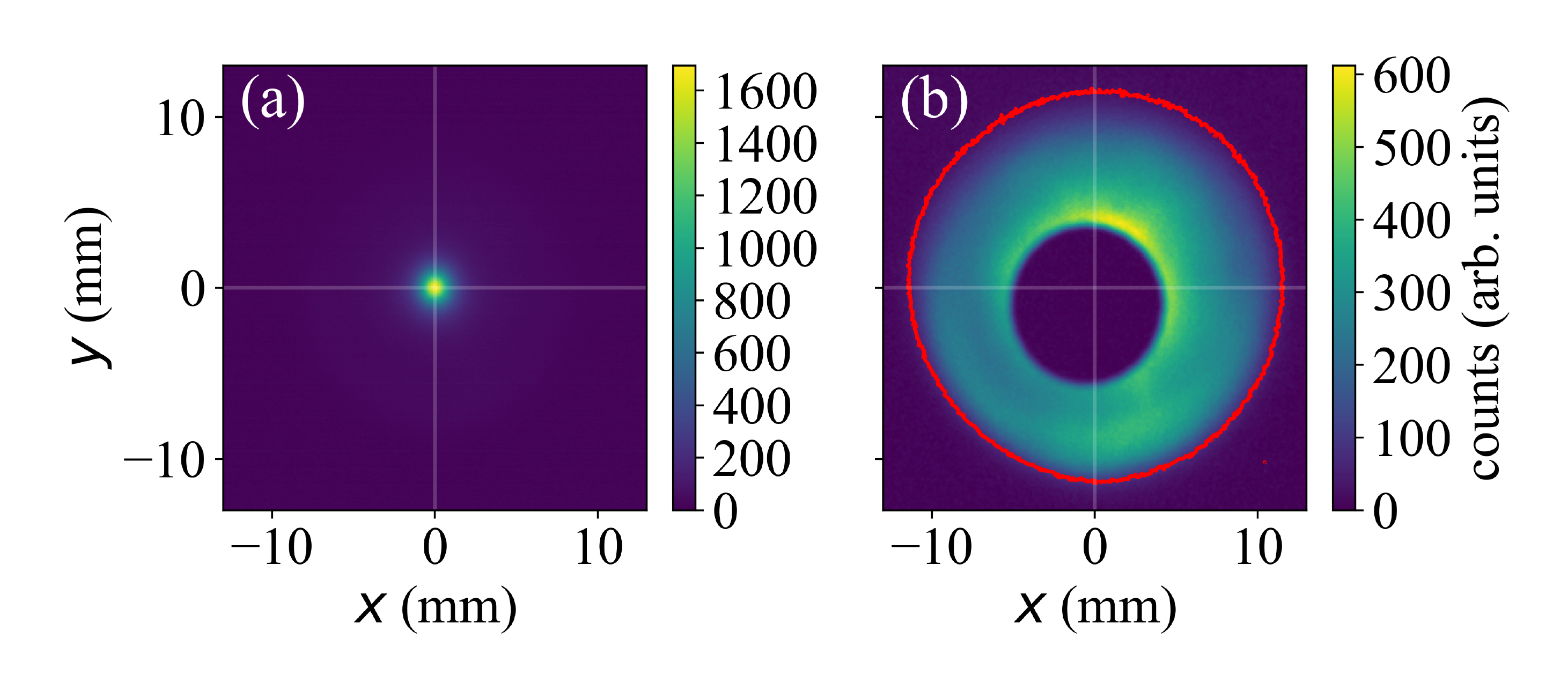}
\caption{Images of beam core (a) and halo (b) at the second (downstream) imaging station. The red solid line in (b) shows the detected halo boundary used for determination of the maximum deflection radius. Note that the intense light emitted by the core of the beam is blocked by a mask.}\label{fig2-images}
\end{figure}

\begin{figure}[tb]
\includegraphics{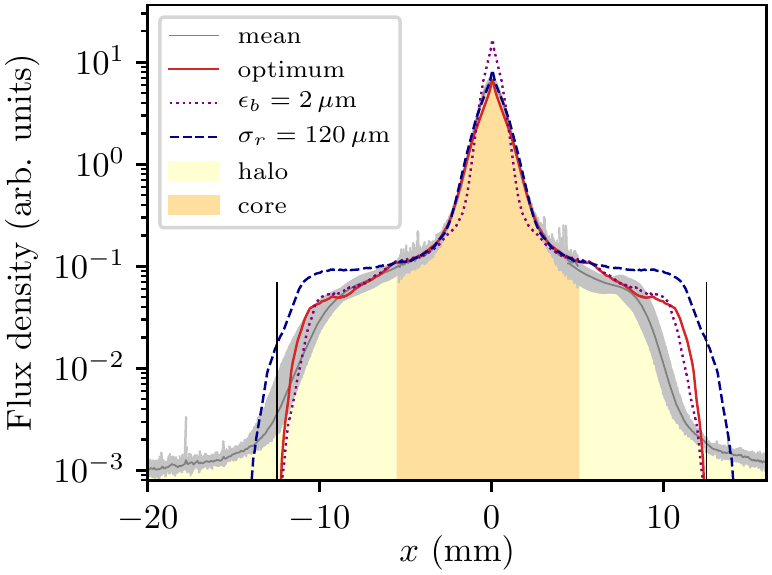}
\caption{Angle-averaged radial distributions of the beam flux in several individual shots (grey lines), additionally averaged over 30 shots (thin dark gray line), and simulated: with optimum parameters from Table~\ref{t1} (solid red line), with a reduced angular spread and the same beam radius (dotted line) and with a reduced beam radius and the same angular spread (dashed line). The vertical black lines mark the maximum proton radius calculated with the contour method \cite{NIMA-909-123}.}\label{fig3-profile}
\end{figure}

The defocused protons form a halo around the micro-bunched part of the beam. Its time-integrated flux is measured at two imaging stations located 1.66\,m and 9.75\,m downstream the plasma cell exit \cite{IPAC17-1682,NIMA-854-100} (Fig.\,\ref{fig1}). Each imaging station separately measures the flux distribution in the halo and in the central part (core) of the beam  with two cameras of different sensitivity (Fig.\,\ref{fig2-images}). Separate registration of the core and halo, followed by stitching two parts of the image, provides a wide dynamic range of measurements (approximately $10^4$) and absolute flux calibration through equating the integral flux and the known beam charge (Fig.\,\ref{fig3-profile}). The halo images processed with a contour method \cite{NIMA-909-123} yield the maximum radius of defocused protons, which is a convenient scalar to be compared with simulations.

The simulations are performed with the quasistatic two-dimensional (axisymmetric) particle-in-cell code LCODE \cite{PRST-AB6-061301,NIMA-829-350}. The simulation window encloses the whole beam and moves with the speed of light $c$ in the positive $z$ direction. The radial window size is always $30c/\omega_p$, where $\omega_p = \sqrt{4 \pi n_0 e^2/m}$ is the plasma frequency, $n_0$ is the plasma density, $e$ is the elementary charge, and $m$ is the electron mass. The grid cell has equal sizes $0.01c/\omega_p$ in both longitudinal and radial directions. There are 10~radius-weighted plasma macro-particles of each species (electrons and ions) in each cell and about $10^6$ or more (proportionally to $\omega_p$) equal beam macro-particles in total. The time step for the beam is $100\omega_p^{-1}$, and the plasma state is updated every $200c/\omega_p$. The plasma has sharp boundaries in both longitudinal and radial directions, as simulations \cite{JPD51-025203,NIMA-740-197,PRA99-063423} have suggested. After the plasma cell, the protons propagate to the screens along straight lines.

\begin{table}[b]
 \caption{Beam and plasma parameters taken as input for simulations. Parameters in the first group are the same in all runs. Parameters in the second group (below the line) may vary in different regimes, and the listed values are those for illustrative cases (Figs.\,\ref{fig3-profile}, \ref{fig5-wall}, \ref{fig6-expl}).}\label{t1}
 \begin{indented}
 \item[]\begin{tabular}{@{}ll}\br
  Parameter, notation & Value \\ \mr
  Plasma length, $L$ & 10.3\,m \\
  Plasma radius, $R_p$ & 1.5\,mm \\
  Plasma ion-to-electron mass ratio, $M_i$ & 157\,000\\
  Plasma initial temperature, $T_e$ & 0\,eV \\
  Beam energy, $W_b$ & 400\,GeV \\
  Beam energy spread, $\delta W_b$ & 0.035\,\% \\
  \mr
  Plasma density, $n_0$ \quad & $1.965\times 10^{14}\,\text{cm}^{-3}$ \\
  Beam population, $N_b$ & $2.86\times 10^{11}$ \\
  Beam length, $\sigma_z$ & 6.57\,cm \\
  Beam radius, $\sigma_r$ & $200\,\mu$m \\
  Normalized beam emittance, $\epsilon_b$ & 3.6\,mm\,mrad \\
  Seed pulse position, $\xi_s$ & 3.75\,cm \\
  \br
 \end{tabular}
 \end{indented}
\end{table}

\begin{figure}[tb]
\includegraphics{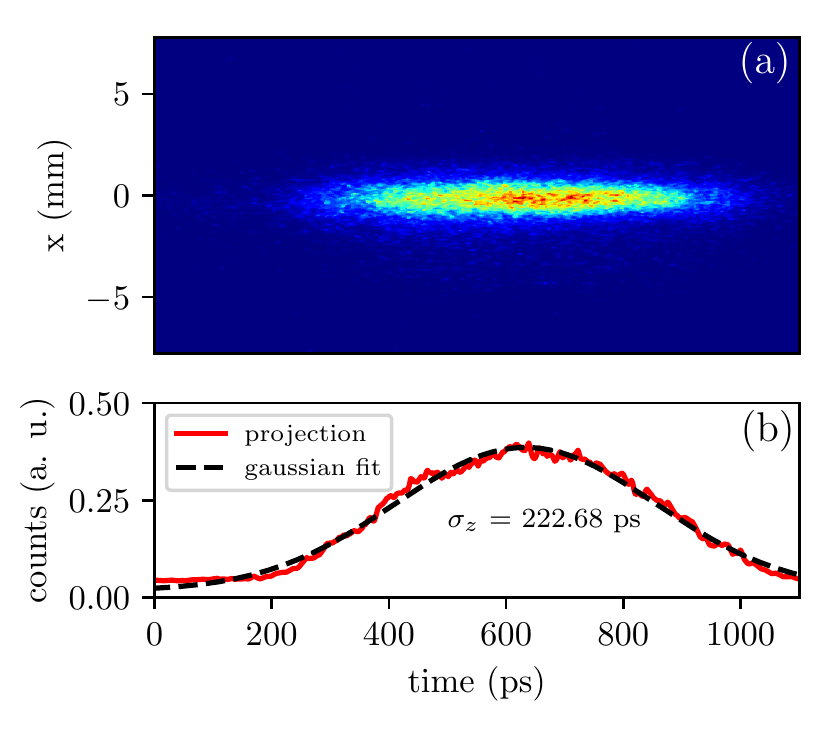}
\caption{Streak camera image of the proton beam (a) and Gaussian approximation of the beam current (b).}\label{fig4-streak}
\end{figure}

Most of the input parameters for simulations (Table~\ref{t1}) are known with sufficient precision and their uncertainties have little effect on quantitative characteristics of beam self-modulation. Less well known are the beam length, radius and emittance and these are determined from the data used in this analysis. The proton beam was modelled with Gaussian distributions in both the longitudinal and transverse directions. The standard deviation of the proton beam in the longitudinal direction, $\sigma_z$, was determined prior to the start of data taking using a streak camera \cite{RSI88-025110}  (Fig.\,\ref{fig4-streak}) and correlated to parameters measured with beam quality monitors in the SPS \cite{sz_at_SPS}. Measurements performed in the SPS during data taking were then used to calculate the bunch length on a shot-by-shot basis.
 


The initial beam radius $\sigma_r$ and emittance $\epsilon_b$ were not measured with the required precision during the discussed parameter scans, so their input values are fitted to the experimental data in four steps. First, we determine the radial beam flux distributions by separate averaging of left and right halves of the images (Fig.\,\ref{fig3-profile}). This allows evaluation of the symmetry of the distribution and verifies that the axisymmetric model is applicable. The systematic asymmetry that we see in the halo is relatively low and can be caused by transverse misalignments between the proton beam and the plasma column. Second, we average the measured distributions over 30 shots. Third, in simulations, we vary the beam angular spread $\delta \alpha \approx \epsilon_b / (\gamma \sigma_r)$, where $\gamma$ is the relativistic factor of the beam. It determines the width of the beam head that propagates in the neutral gas and therefore the shape of the core part of beam flux distribution (compare red solid and magenta dotted lines in Fig.\,\ref{fig3-profile}). Fourth, we adjust the beam radius to match the height and width of the flux profile ``shoulders'' (compare red solid and blue dashed lines in Fig.\,\ref{fig3-profile}). The height and width of the shoulders depend on the strength of the fields produced in the plasma and are therefore sensitive to the initial beam density, which scales as $\sigma_r^{-2}$. Even with the best parameter fit (red line), the simulated beam profile is sharply peaked at the axis and therefore locally differs from the measured profile. This happens because both radial and angular distributions of beam micro-bunches in the plasma are strongly peaked in the perfectly axisymmetric case \cite{PoP24-023119}, while in the experiment this narrow spike is blurred. The obtained values of $\sigma_r$ and $\epsilon_b$ depend weakly on beam population. We approximate them as
\begin{equation}\label{em}
    \epsilon_b [\mu\mathrm{m}] = 2.3 + 4.5 \times 10^{-12} N_b, \ \ \sigma_r [\mu\mathrm{m}] = 105 \sqrt{\epsilon_b [\mu\mathrm{m}]},
\end{equation}
so the beam parameters vary from $\epsilon_b = 2.75$\,mm\,mrad and $\sigma_r = 175\,\mu$m at $N_b = 10^{11}$ to the values listed in Table~\ref{t1} at higher $N_b$.

The simulated maximum radius of deflected protons $r_\text{max}$ depends on the number of beam macro-particles used. The larger the number, the better we resolve the ``wings'' of the Gaussian angular distribution, the larger the maximum initial transverse momentum beam macro-particles can have, and the larger the maximum deflection we observe. To avoid this ambiguity, we consider $r_\text{max}$ as the radius at which the simulated beam flux density equals the noise level in experiments.

\begin{figure}[tb]
\includegraphics{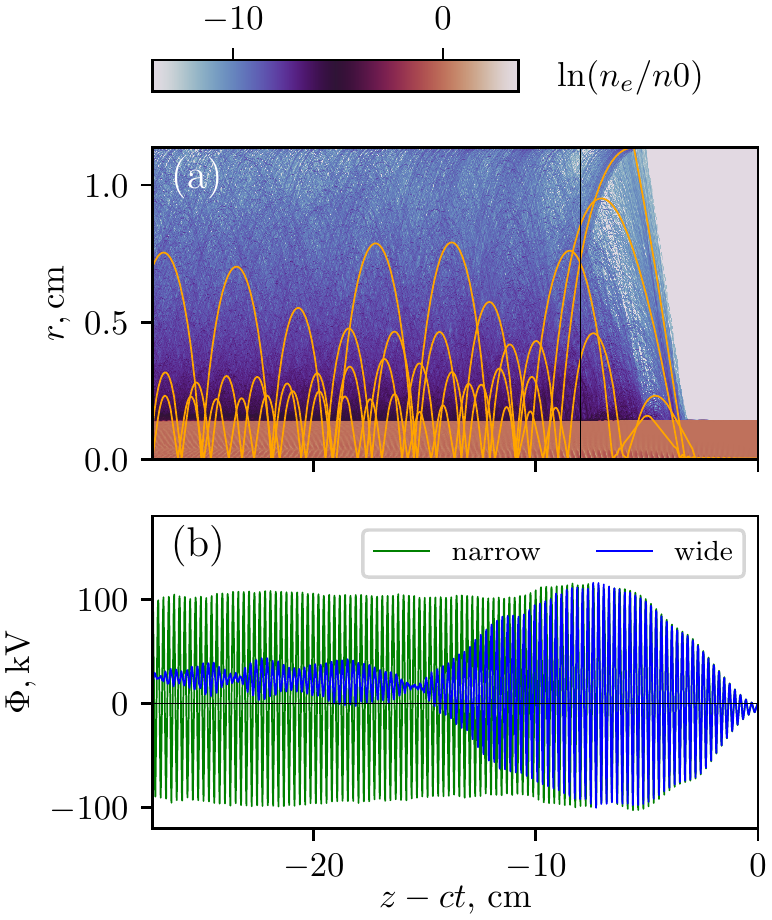}
\caption{(a) Density of plasma electrons (color map) and trajectories of individual electrons (lines) after 4 meters of beam propagation in the plasma. The vertical black line shows the cross-section detailed in Fig.\,\ref{fig6-expl}. (b) Wakefield potential $\Phi$ on the axis after the same propagation distance calculated with wide ($30c/\omega_p \approx 11.4\,\text{mm}$) and narrow ($1.7\,\text{mm} \approx 4.5c/\omega_p$) simulation windows.}\label{fig5-wall}
\end{figure}

\begin{figure}[tb]
\includegraphics{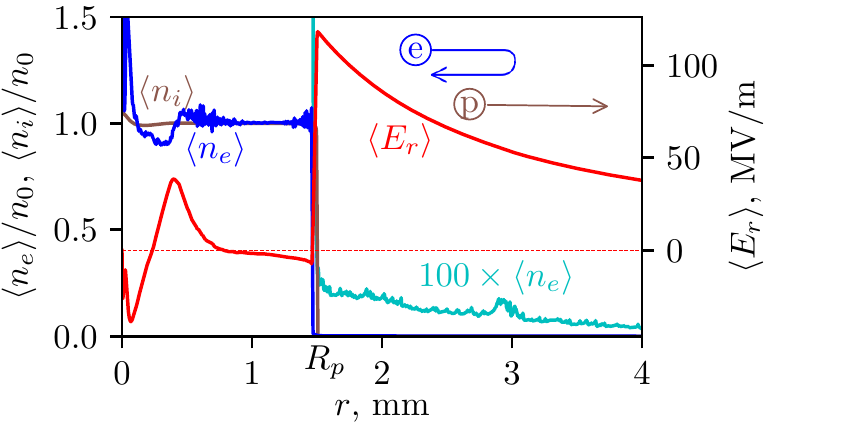}
\caption{Radial dependencies of plasma-period-averaged ion density $\langle n_i \rangle$, electron density $\langle n_e \rangle$, and radial electric field $\langle E_r \rangle$ 8\,cm behind the seed laser pulse after 4 meters of beam propagation in the plasma. The electron density outside the plasma column is 100 times magnified for visibility. Arrows schematically show the field effect on plasma electrons and beam protons.}\label{fig6-expl}
\end{figure}

\begin{figure}[tb]
\includegraphics{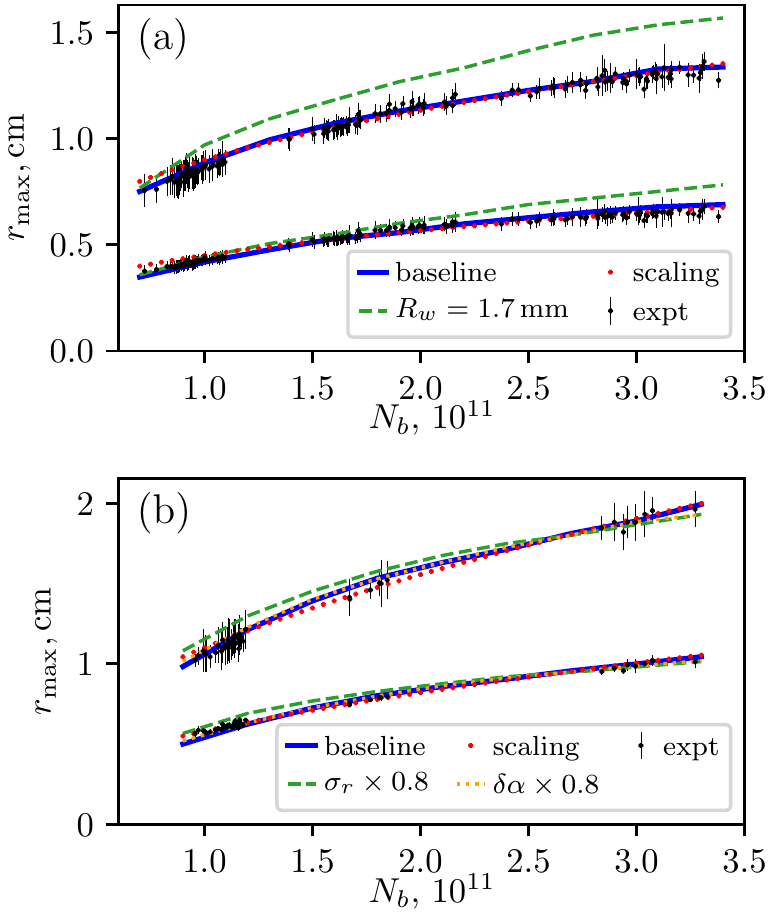}
\caption{Beam radius on the screens for low (a) and high (b) density plasmas and varying beam population. Top and bottom groups of lines in each panel are for two imaging stations. Black dots with error bars (``expt'') are experimentally measured values (individual shots), lines are simulation results. The error bars were determined the same way as in \cite{NIMA-909-123} and show the standard deviation of the determined radii along different directions of the image. Solid blue lines are for optimum parameter sets. The green dashed line in (a) shows the effect of a narrow simulation window ($R_w = 1.7$\,mm). The additional lines in (b) are for narrower beam (with $\sigma_r$ reduced by 20\% at the same $\delta \alpha$, green dashed line) or lower-emittance beam ($\delta \alpha$ reduced by 20\% at the same $\sigma_r$, orange dotted line). Dotted red lines are scalings $\propto N_b^{1/3}$ in (a) and $\propto N_b^{1/2}$ in (b).}\label{fig7-charge}
\end{figure}

The wide simulation window ($30c/\omega_p$) is necessary for correct calculation of high-energy electron trajectories. These electrons appear as a result of wavebreaking, which often occurs in the simulated regimes after beam micro-bunching [Fig.\,\ref{fig5-wall}(a)]. The electrons are ejected from the plasma column radially and leave an unbalanced positive charge behind \cite{PRL112-194801, PoP25-063108}. The charge separation electric field keeps them from traveling far, so these electrons form a negatively charged halo around the positively charged plasma column (Fig.\,\ref{fig6-expl}). The halo electrons move predominantly in the positive $z$ direction \cite{PRL112-194801} and efficiently interact with the wakefield when crossing the near-axis area: they can take energy from some regions or release it in others. As a consequence, the wakefield damps when a substantial number of halo electrons returns to the near-axis area. This is seen when comparing trajectories of halo electrons [Fig.\,\ref{fig5-wall}(a)] and the wakefield potential on the axis [Fig.\,\ref{fig5-wall}(b)]. The wakefield potential $\Phi$ characterizes the force exerted on axially moving ultra-relativistic particles:
\begin{equation}\label{e1}
    E_z = - \frac{\partial \Phi}{\partial z}, \qquad E_r - B_\varphi = - \frac{\partial \Phi}{\partial r}.
\end{equation}
A reduced amplitude of potential oscillations on the axis evidences reduction of both accelerating and defocusing properties of the wakefield.

\section{Comparison of simulations and experiments}

We compare simulations and experiments in three parameter scans. The first two are beam population scans at low ($1.965\times 10^{14}\,\text{cm}^{-3}$) and high ($7.7\times 10^{14}\,\text{cm}^{-3}$) plasma densities (Fig.\,\ref{fig7-charge}). The best agreement variants (blue lines) correspond to the approximation (\ref{em}) and other parameters from Table~\ref{t1}, with the exception that the high density scan was with $\sigma_z = 7$\,cm and $\xi_s = 1.9$\,cm.

The green dashed line in Fig.\,\ref{fig7-charge}(a) shows the effect of a narrow simulation window. Disagreement with experiments is expected, but the sign of the effect may seem counter-intuitive. If the simulation wall is located close to the plasma boundary, the positive electric field around the plasma (Fig.\,\ref{fig6-expl}) does not additionally accelerate beam protons in the radial direction. However, the narrow window also ``switches off'' the effect of halo electrons, the wakefield becomes stronger, and we observe larger proton deflection in simulations.

The additional lines in Fig.\,\ref{fig7-charge}(b) illustrate the sensitivity to several beam parameters.  The simulations agree well with the measured data over most of the range of proton beam population for variations of 20~\% in the beam radius and the beam angular divergence. The largest discrepancy comes from varying the beam radius at small bunch populations.  In this case, a 20\% reduction in the beam radius results in visible disagreement with measurements (green dashed line) for the second imaging station. The sensitivity to the angular spread is lower: a 20\% reduction changes $r_{\rm max}$ almost imperceptibly (orange dotted line). Variation of other beam parameters also have small effects on proton defocusing, except the seed position discussed in Ref.\,\cite{Marlene}.

Theory \cite{PoP20-083119} suggests that the maximum wakefield at AWAKE may be limited by nonlinear elongation of the plasma wave period. The wave goes out of resonance with the bunch train, and electric field growth saturates at about $0.4 mc\omega_p/e$. If this mechanism is indeed the main limiting factor, then the maximum wakefield amplitude should scale as the contribution of an individual bunch to the power of 1/3, or as $N_b^{1/3}$ (see Eq.\,(12) of \cite{PoP20-083119}). The dependence of maximum proton deflection on beam population at the low plasma density follows this scaling remarkably well [dotted line in Fig.\,\ref{fig7-charge}(a)]. At the high density, however, the maximum radius scales as $N_b^{1/2}$ [dotted line in Fig.\,\ref{fig7-charge}(b)], which indicates that the microbunch train is not dense or long enough to excite the nonlinear wakefields at this plasma density. In this case, the radial momentum gained by the deflected protons $p_\text{max}$ is defined by the depth of the wakefield potential well $e\Phi_\text{max}$ that scales as $N_b$. As a result, $p_\text{max} \propto \sqrt{2 e \Phi_\text{max} W_b}/c \propto \sqrt{N_b}$.

\begin{figure}[htb]
\includegraphics{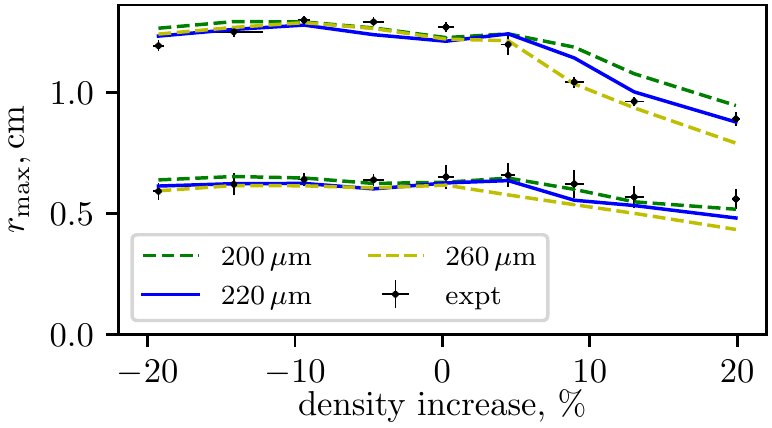}
\caption{Beam radius on the screens for different plasma density gradients (characterized by the relative density increase at the downstream plasma end): black dots are the experimentally measured values; error bars show the standard deviation of all individual measurements (about 30 shots) at a given gradient setting, lines are simulation results for different beam radii.}\label{fig8-grad}
\end{figure}

For the density gradient scan, the simulations also follow the measured trend (Fig.\,\ref{fig8-grad}). Here the plasma density changes linearly along the cell due to a temperature difference between the two Rb reservoirs attached to the cell at the ends \cite{JPD51-025203}. The density at the upstream end is always $1.8\times 10^{14}\,\text{cm}^{-3}$. Beam parameters are those from Table~\ref{t1}, except $\sigma_z = 7.4$\,cm and $\xi_s = 2.4$\,cm. 
The simulations show significant sensitivity to the beam radius for positive gradients larger than $+5$\%.  Overall, the best agreement is for $\sigma_r \approx 220\,\mu$m.

\begin{figure}[htb]
\includegraphics{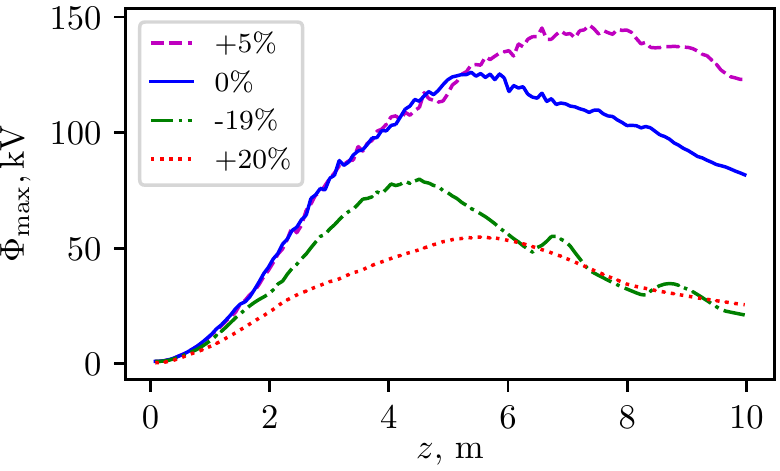}
\caption{Maximum amplitude of the wakefield potential $\Phi_{\rm max}$ versus beam propagation distance $z$ for $\sigma_r = 220\,\mu$m and different magnitudes of relative density increase along the plasma cell.}\label{fig9-gmaxf}
\end{figure}
Small positive gradients contribute to beam self-modulation by increasing the total charge of microbunches \cite{NIMA-829-63}. The effect is similar to that of the density step \cite{PoP18-024501,PoP22-103110}: larger fractions of initially formed micro-bunches remain focused by the wave because of plasma wavelength shortening. With small positive gradients (density increase about 2\% over 10 meters), the longitudinal wakefield is stronger \cite{NIMA-829-63}, and externally injected electrons gain a higher energy \cite{Nat.561-363}. Higher positive gradients, however, are destructive and reduce the maximum wakefield (Fig.\,\ref{fig9-gmaxf}), which results in weaker deflection of defocused protons (Fig.\,\ref{fig8-grad}). High negative gradients also reduce the wakefield, but not the maximum deflection radius. This happens because the strongest defocused protons do not always experience the strongest wakefield generated by the beam. Negative density gradients change the wavelength so that the micro-bunches mostly fall into the defocusing wave phase, and more efficient coupling with the transverse field compensates the lower wakefield amplitude.

\section{Summary}

To conclude, simulations with two-dimensional (axisymmetric) particle-in-cell code LCODE agree with the measurements of the maximum radius of deflected protons in three different experimental scans: proton beam population scans ($1 \div 3 \times 10^{11}$ particles) for high and low plasma densities, and gradient scan ($-20 \div 20\%$). The difference between calculated and measured values is about or below 5\%. This means that particle-in-cell codes can be used to interpret the physics underlying experimental data on SSM. However, the agreement with experiments can be achieved only for wide enough simulation windows ($\sim 30c/\omega_p$ for AWAKE parameters) that are necessary to correctly account for ejected plasma electrons that charge the plasma column positively and then return to the axis destroying the wakefield. This additionally increases the computational power required for simulations of long microbunch trains in a plasma.

\ack

This work was supported in parts by the Foundation for the Development of Theoretical Physics and Mathematics ``BASIS''; a Leverhulme Trust Research Project Grant RPG-2017-143 and by STFC (AWAKE-UK, Cockcroft Institute core, John Adams Institute core, and UCL consolidated grants), United Kingdom; a Deutsche Forschungsgemeinschaft project grant PU 213-6/1 ``Three-dimensional quasi-static simulations of beam self-modulation for plasma wakefield acceleration''; the National Research Foundation of Korea (Nos.\ NRF-2016R1A5A1013277 and NRF-2019R1F1A1062377); the Portuguese FCT---Foundation for Science and Technology, through grants CERN/FIS-TEC/0032/2017, PTDC-FIS-PLA-2940-2014, UID/FIS/50010/2013 and SFRH/IF/01635/2015; NSERC and CNRC for TRIUMF's contribution; the U.S.\ National Science Foundation under grant PHY-1903316; the Wolfgang Gentner Programme of the German Federal Ministry of Education and Research (grant no.\ 05E15CHA); and the Research Council of Norway. A.A.\,Gorn and K.V.\,Lotov acknowledge the support of the Russian Foundation for Basic Research (grant 19-32-90125). M. Wing acknowledges the support of the Alexander von Humboldt Stiftung and DESY, Hamburg. Support of the Wigner Datacenter Cloud facility through the
``Awakelaser'' project and the support of P\'eter L\'evai is acknowledged. The work of V. Hafych has been supported by the European Union's Framework Programme for Research and Innovation Horizon 2020 (2014--2020) under the Marie Sklodowska-Curie Grant Agreement No.\ 765710.  The AWAKE collaboration acknowledge the SPS team for their excellent proton delivery. LCODE simulations were performed on HPC-cluster ``Akademik V.M. Matrosov'' \cite{matrosov}.

\end{document}